\documentclass[review,12pt]{elsarticle}

\makeatletter
\def\ps@pprintTitle{}%
\makeatother

\usepackage{graphicx}
 \usepackage{geometry}
\usepackage{amsfonts}
\usepackage{amsmath}
\usepackage{amssymb}
\usepackage{amstext}
\usepackage{epstopdf}
\usepackage{makeidx}

\usepackage{amsthm}

\usepackage{graphicx,color}
\makeindex

\begin{document}
\title{Hybrid Monte Carlo Simulation of Stress-Induced Texture Evolution with Inelastic Effects}
\author{Liangzhe Zhang$^{a,b,}$\footnote{Corresponding author. Tel.: 718-260-3082; fax: 718-260-3532. E-mail: lzhang@poly.edu (L. Zhang)}, R\'emi Dingreville$^{a}$, Timothy Bartel$^{c}$, Mark T. Lusk$^{b}$}
\address{$^{a}$Department of Mechanical and Aerospace Engineering, NYU-Poly, Brooklyn, NY 11201\\
$^{b}$ Department of Physics, Colorado School of Mines, Golden, CO 80401 \\
$^{c}$Sandia National Laboratories, Albuquerque, NM 87185}

\begin{abstract}
A hybrid Monte Carlo (HMC) approach is employed to quantify the influence of inelastic deformation on the microstructural evolution of polycrystalline materials. This approach couples a time explicit material point method (MPM) for deformation with a calibrated Monte Carlo model for grain boundary motion. A rate-independent crystal plasticity model is implemented to account for localized plastic deformations in polycrystals. The dislocation energy difference between grains provides an additional driving force for texture evolution. This plastic driving force is then brought into a MC paradigm via parametric links between MC and sharp-interface (SI) kinetic models. The MC algorithm is implemented in a parallelized setting using a checkerboard updating scheme. As expected, plastic loading favors texture evolution for grains which have a bigger Schmid factor with respect to the loading direction, and these are the grains most easily removed by grain boundary motion. A macroscopic equation is developed to predict such texture evolution.
   
{\bf $Keywords:$} Monte Carlo, grain boundary, plasticity, anisotropy, texture, driving force
\end{abstract}
\maketitle

\section{Introduction}
The prediction of microstructural evolution in response to thermo-mechanical loading is important for materials design, processing or thermomechanical fatigue phenomena. Computational modeling of evolving texture in response to large plastic deformation and recrystallization has been studied extensively~\cite{Bronkhorst.1992,Dillarmore.1974,Hirsch.1988.02,Roters20101152} but less so than that produced by thermally-induced stresses--i.e. stress-induced texture evolution~\cite{Telang.2007,Battaile.1999}. We consider a thermo-mechanical setting in which temperature changes cause stresses to develop due to geometrical constraints. The temperature is sufficiently high to generate grain boundary motion and yet low enough such that recrystallization does not occur. The induced stresses may be associated with both elastic and plastic deformation~\cite{Telang.2007}. 

In a previous work, a Hybrid Monte Carlo (HMC) approach~\cite{Zhang2010419} was developed by combining a MC algorithm for grain boundary motion~\cite{Laudau.2005} with the Material Point Method (MPM)~\cite{Sulsky.1995} for elastic deformation. Purely elastic driving forces, originating from the anisotropic mechanical response of individual grains, are treated as a bulk body force in a Potts model for grain boundary  evolution~\cite{Wu.1982}. The approach is time accurate through the use of parametric links~\cite{Laudau.2005} to sharp-interface (SI) kinetics~\cite{Abeyaratne.1990}. It also takes advantage of the fact that MC grain boundary mobility is independent of the driving force~\cite{Liu.2002,Zhang2010790,Lobkovsky.2004}. 

The present work extends this paradigm to include the influence of inelastic deformation on texture evolution~\cite{ Battaile.1999}. As in the elastic study~\cite{Zhang2010419}, texture evolution is assumed to be dominated by grain boundary kinetics~\cite{Gottstein20051535,Gottstein20061065}. Furthermore, we consider infinitesimal deformation to distinguish the stress-induced texture from deformation texture. The latter is associated with grain and lattice rotation in response to finite deformations~\cite{Dillamore.1964,Kocks.1998}. A stochastic, crystal plasticity model, developed from rate-independent crystal plasticity~\cite{Miehe.2001}, is applied within the MPM framework as the constitutive model to capture the elasto-plastic response of a polycrystalline media~\cite{Dingreville2010617}. As opposed to conventional deterministic algorithms~\cite{Miehe.2001}, the stochastic algorithm relies on a MC routine to determine the activated slip system which is therefore referred to as the Monte Carlo Plasticity (MCP). When plastic deformation occurs, dislocations are generated, stored and annihilated within the microstructure. The heterogeneous distribution of these dislocations within the polycrystalline medium constitutes a plastic driving force for grain boundary migration. This is treated as a body force within the MC kinetics using parametric links between MC and SI models. A Red/Black (RB) updating scheme is used to parallelize the MC algorithm~\cite{Fried.1990}, although other methods might also be useful~\cite{Bortz.1975}.  This parallelized HMC approach is used to investigate the microstructural evolution of nickel polycrystals under plastic loading. As expected, the grains with smaller Schmid factors gradually dominate the polycrystalline system. The data is subsequently used to construct a macroscopic kinetic equation to predict the evolution of microstructure.

\section{Monte Carlo crystal plasticity}
Plastic response of polycrystalline materials is treated through a classical rate-independent small deformation crystal plasticity formulation~\cite{Havner.1969}. The foundations of the constitutive model assume that the elasto-plastic response of single crystals is dominated by slip deformation mechanisms~\cite{Taylor.1938,Havner.1982,Beaudoin20003409}. A successful numerical algorithm must carry out three tasks: the determination of activated slip systems; the calculation of the plastic slip on each activated slip system; and, the solution of redundant constraints associated with a hardening law~\cite{Miehe.2001,Anand.1996}. Various numerical methods have been devised and successfully implemented in deterministic formats~\cite{Miehe.2001,Anand.1996}.  

As opposed to deterministic algorithms, the current work adopts a probabilistic approach borrowed from concepts in statistical mechanics in which only one slip system is activated during each time step. Plastic slip is therefore treated as a series of discrete, probabilistic events that mimic the sequential accumulation of dislocations at the lattice scale. This Monte Carlo crystal plasticity (MCP) is algorithmically simple because plastic slip can be resolved through the solution of one equation with no redundant constraints. On the other hand, the associated computational steps has to be sufficiently small such that a sequence of single slips mimics multiple slip behavior. A probabilistic algorithm, detailed in what follows, is used to determine which slip system is chosen at each step. The constitutive framework and stress updating routine are otherwise standard~\cite{Miehe.2001}.

Given a set of potentially activated slip systems $\zeta=(1,2,...,n)$, identified through comparison of resolved shear stress with slip resistance~\cite{Miehe.2001,Counts.2008}, the elastic energy of a crystal, $E_{\alpha}$, can be calculated if each slip system of the set $\zeta$ is individually activated. This generates $n$ possible states for the deformed crystal. The probability, $p_{\alpha}$, of a slip system being selected is computed using the partition function~\cite{Pathria.1996}, $Z$ :
\begin{equation}
\begin{tabular}
[c]{l}$Z=\sum\limits_{\alpha=1}^{n}e^{-\beta E_{\alpha}}$\\$p_{\alpha}=\frac{e^{-\beta E_{\alpha}}}{Z}$,
\end{tabular}
\label{MC_pla}
\end{equation} 
where $\alpha$ is the index of a potentially activated slip system, and $\beta$ is the inverse of the fundamental temperature. Dislocation energy can be ignored in Eqn. (\ref{MC_pla}) due to the fact that an isotropic hardening model is used~\cite{Miehe.2001,Counts.2008}. The $ith$ slip system of the set $\zeta$ is activated when the following criterion is met:
\begin{equation}
\begin{tabular}
[c]{l}$\sum\limits_{\alpha=1}^{i}p_{\alpha}-p_{i}\leq R<\sum\limits_{\alpha=1}^{i}p_{\alpha}$.
\end{tabular}
\label{random}
\end{equation}
Here $R$ is taken as a random number between $0$ and $1$. With the activated slip system determined, the deformation can easily be parsed into elastic and plastic components.

\begin{figure}[ptb]\begin{center}
\includegraphics[width=0.45\textwidth]{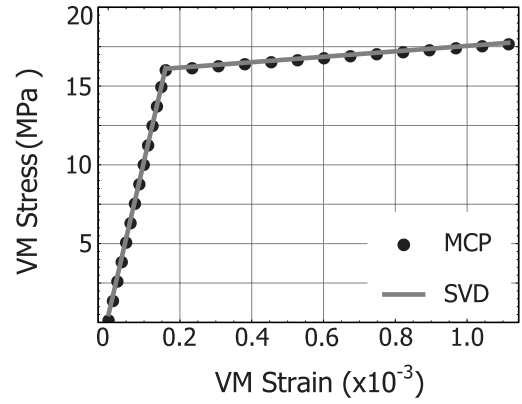}
\caption{Comparison of stress-strain relationship between MCP and SVD algorithms with a prescribed strain increment. The discrete dots and solid line represent MCP and SVD results, respectively.}
\label{stress_mcp_svd}
\end{center}\end{figure}

As a verification of this MCP method, a uniaxial tension test was carried out on a single crystal of nickel. The material properties~\cite{Miehe.2001,Counts.2008} required for this model are listed in Table \ref{plapara}.
\begin{table}[table1] \centering
\begin{tabular}{|l|l|l|}
\hline
Elastic moduli & C$_{11}$ & \ \ \ 250.8 GPa \\ \hline
Elastic moduli & C$_{22}$ & \ \ \ 150.0 GPa \\ \hline
Elastic moduli & C$_{44}$ & \ \ \ 123.5 GPa \\ \hline
CRSS & $\tau _{0}$ & \ \ \ 8 MPa \\ \hline
Saturation stress & $\tau _{s}$ & \ \ \ 0.122 GPa \\ \hline
Initial hardening & h$_{0}$ & \ \ \ 0.44 GPa \\ \hline
Inverse fundamental temp & $\beta$ & \ \ \ 1000.0 \\ \hline
\end{tabular}
\caption{Nickel properties required in MCP model~\cite{Miehe.2001,Counts.2008}.
\label{plapara}}
\end{table}
A prescribed strain increment, $\Delta\varepsilon _{x},$ was enforced at each time step:
\begin{equation}
\Delta \varepsilon _{x}=\left[ 
\begin{tabular}{lll}
$\epsilon $ & $\ 0$ & $\ 0$ \\ 
$0$ & $-\frac{1}{2}\epsilon $ & $\ 0$ \\ 
$0$ & $\ 0$ & $-\frac{1}{2}\epsilon $%
\end{tabular}%
\ \ \right] ,
\end{equation}
with $\epsilon =1.0\times 10^{-6}.$ Fig. \ref{stress_mcp_svd} compares the results obtained using the MCP logic with their counterparts obtained with a deterministic model wherein a singular value decomposition (SVD) algorithm is applied to solve the ill-conditioned constraint equations. The two approaches clearly agree. 

\section{Grain boundary kinetics}
\begin{figure}[ptb]\begin{center}
\includegraphics[width=0.45\textwidth]{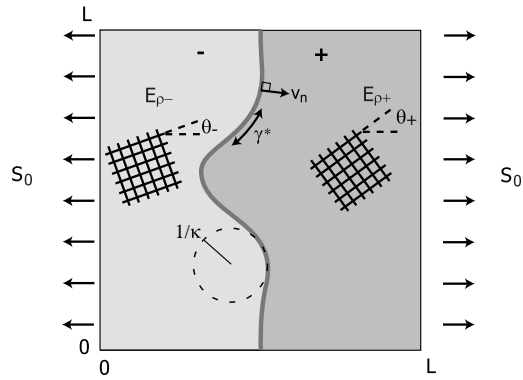}
\caption{A single grain boundary with SI notation.}
\label{si_schem}
\end{center}\end{figure}

Sharp-interface, grain boundary kinetics~\cite{Abeyaratne.1990} is the basis for our grain boundary kinetics model. It can be developed by considering a bi-crystal subjected to uniaxial loading that results in plastic deformation as shown in Fig. {\ref{si_schem}}. 
The dislocation energy, $E_{\rho}$, and dislocation density, $\rho$ are related using the relation~\cite{Humphreys.2004}
\begin{equation}
E_{\rho}= \frac{1}{2}\rho Gb_{g}^{2}  , \quad
\rho = (\frac{2\sigma_{f}}{Gb_{g}})^{2}  .
\label{pla_dri}
\end{equation}
Here $G$ is the shear modulus, $b_{g}$ is the Burgers vector, and $\sigma_{f}$ is the flow stress. The thermodynamic driving force for interfacial accretion is given by~\cite{Abeyaratne.1990}
\begin{equation}
f_{s}=[\![E_{\rho}]\!]+\kappa \gamma^{*} \ ,
\label{driv_trac}
\end{equation}
where $\gamma^{\ast} $ is the capillary driving force, $\kappa $ is the local mean curvature, and $[\![\symbol{126}]\!]$ is the jump in a field across the grain boundary. The elastic driving force is temporarily suppressed in order to study the influence of the plastic behavior on texture evolution. The Herring relation~\cite{Herring.1949} is used to describe the accretive normal speed of the interface, $v_{n}$:
\begin{equation}
v_{n}=m_{s}f_{s}\ ,  \label{si_kins_1}
\end{equation}
where $m_{s}$ is grain boundary mobility.

The MC paradigm is intended to implement SI kinetics within a computationally efficient setting. Continuum, deterministic fields are linked to discrete, probabilistic counterparts, and the physical domain is discretized into a square lattice. A Q-state Potts model~\cite{Wu.1982} is used to represent a crystalline system and grain orientation is described by an integer-valued spin field, $q_{i}$, which ranges from $1$ to $Q$. The system Hamiltonian is then written as
\begin{equation}
\mathcal{H}=\sum_{i=1}^{N}\sum_{n=1}^{M}J_{q_{i},q_{n}}\delta _{q_{i},q_{n}}+\sum_{i=1}^{N}b_{i},
\label{H_1}
\end{equation}
where $b_{i}$ is the spatially varying bulk energy associated with the $i^{th}$ lattice, $J_{q_{i},q_{n}}$ is the interaction energy between the neighboring spin fields $q_{i}$ and $q_{n}$, $M$ is the number of neighbors considered on the selected lattice, and $\delta$ is the Kronecker delta. Kinetics, within the MC paradigm, is translated to a series of probabilistic trials carried out at all lattice sites. The acceptance of a trial event is determined by a probability function based on the change in the Hamiltonian associated with a trial flip. Plastic effects can therefore be accounted for if the plastic driving force is treated as a bulk energy term in Eqn. (\ref{H_1}). 

Parametric links between SI and MC models are used to convert SI parameters into MC format~\cite{Zhang2010419}:
\begin{align}
\overline{\gamma^{\ast}}_{s}&=\frac{\gamma^{\ast} E_{0}}{ \Delta^{2}J l_{0}^{2}} \notag \\
\overline{b}_{s}& =\frac{2b E_{0}}{\Delta ^{3}l_{0}^{3}}  \notag \\
\overline{m}_{s}& =\frac{\Delta ^{4}m(\alpha_{mc}, \theta)l_{0}^{4}}{J\tau _{mc}E_{0} t_{0}}  ,
\end{align}
where $\overline{\gamma^{\ast}}_{s}$ and $\overline{m}_{s}$ are reasonable grain boundary stiffness and grain boundary mobility obtained from the literature~\cite{Humphreys.2004,Olmsted20093694,Saindrenan.1999}, while $\overline{b}_{s}$ is the computed bulk energy density associated with the SI model. The counterparts in the MC paradigm, which can be analytically or numerically derived, are $\gamma^{\ast}$~\cite{Onsager.1944,Binder.1982,Burkner.1983}, $b$ and $m$~\cite{Liu.2002,Zhang2010790,Zhang2010419,Lobkovsky.2004}, respectively. A characteristic energy, $E_{0}$, length, $l_{0}$, and time, $t_{0}$ along with the MC lattice size $\Delta$, interaction energy $J$, effective temperature $\alpha_{mc}$, inclination angle $\theta$ and time step $\tau _{mc}$, are also required to bridge the SI and MC paradigms~\cite{Zhang2010790}. These links endow the MC simulation with physical time and length scales.

\section{A Parallelized Monte Carlo Algorithm}
In a typical MC implementation, an update of variables immediately follows the acceptance of a trial event as in the Metropolis algorithm for example~\cite{Laudau.2005,Metropolis.1953}. To improve the efficiency of this approach within a parallelized setting, we adopted a Red/Black (RB) updating rule~\cite{Fried.1990} wherein the domain is decomposed into a checkerboard, and each MC step is divided into red/black and black/red half steps. This is an alternative to the N-fold way algorithm~\cite{Bortz.1975,Hassold.1993,Korniss.1999}. In contrast to the large number of updates associated with a standard MC time step, states are not updated in the RB format until each half step finishes. When applied within a parallel environment, the domain is divided uniformly among processors with synchronous communication between processors at each half-step.

\begin{figure}[ptb]\begin{center}
\includegraphics[width=0.6\textwidth]{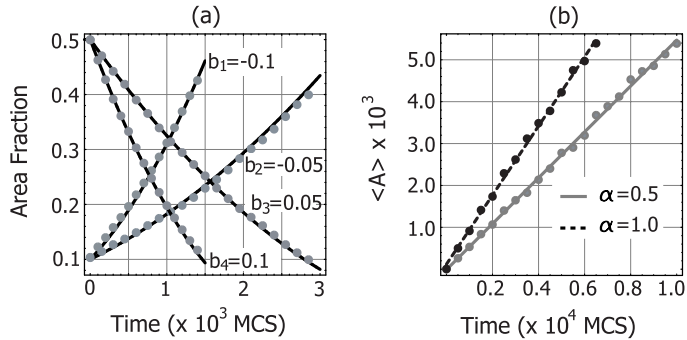}
\caption{Validation of the parallelized MC algorithm in 2D: (a) Comparison of the inner grain area evolution of a bi-crystal between MC and SI results with four combinations of capillary and bulk driving forces at $\alpha_{mc}$=1.4. Solid line and discrete points represent MC and SI results respectively. Grid size is $500\times 500$. (b) Isotropic grain growth at two MC temperatures for 2-D polycrystals. The gray and black dots demonstrate $\alpha_{mc}$=0.5 and $\alpha_{mc}$=1.0, respectively. Grid size is $500\times 500$. MC results are averaged over $10$ samples.}
\label{2D_RB}
\end{center}\end{figure}

The approach is first implemented within a 2D, bi-crystal setting~\cite{Zhang2010790}. A circular grain is placed at the center of a square domain and the inner grain shrinks or grows due to the combined effects of capillary and bulk driving forces. The two-state square lattice Ising model is then employed to represent the physical system. MC simulations were carried out on a $500 \times 500$ grid, at an effective temperature $\alpha_{mc}=1.4$. The evolution of inner grain was simulated with four combinations of capillary and bulk driving forces. Capillary driving force was fixed by giving a constant interaction energy, i.e., $J=1$, while bulk driving force, which is half of the bulk energy difference per site between inner and outer grains, varies from $-0.1$ to $0.1$. In particular, the bulk energy term will be used later on to represent mechanical effect (dislocation energy) in realistic plastic deformation. Corresponding SI simulations were also run using the established parametric links between MC and SI models. Fig. \ref{2D_RB} (a) indicates good agreement between the two models in the prediction of inner grain area as a function of time. An additional test case was also performed for a polycrystalline system where the Q-state ($Q=91$) Potts model with isotropic interaction energy ($J=1$) was used. Many MC simulations were carried out at effective temperatures $\alpha_{mc}=1.0$ and $\alpha_{mc}=0.5$, respectively. Fig. \ref{2D_RB} (b) shows the measured average grain size with linear fits to the data. The linear grain growth behavior is consistent with that of the \emph{Isotropic Grain Growth Theory}~\cite{Holm.2001,Holm.1993}. In later applications, a relatively high MC effective temperature is preferred to remove lattice pinning~\cite{Holm.2001}.

\section{Texture evolution}
\begin{figure}[ptb]\begin{center}
\includegraphics[width=0.6\textwidth]{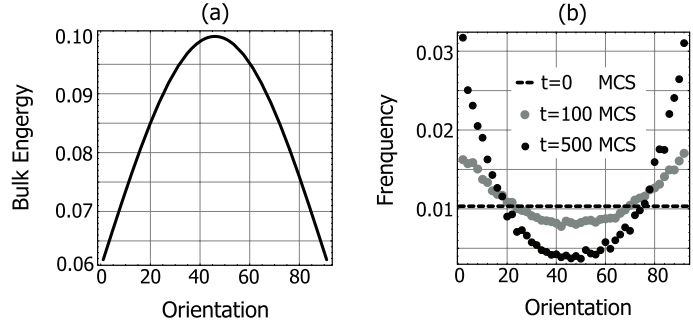}
\caption{The impact of anisotropic bulk energy on polycrystalline texture evolution: (a) Gaussian distribution of orientation-dependent nondimensional bulk energy, and (b) orientation distributions at three time slices. The grid size is $500\times 500$, and $\alpha_{mc}=1.0$. Simulation results were averaged over $100$ runs.}
\label{MC_texture}
\end{center}\end{figure}

The aforementioned parallelized MC algorithm was next applied within the previous polycrystalline setting to study the influence of bulk energy distribution on the evolution of texture. The interaction energy was taken to be isotropic and a Gaussian distribution for the bulk energy was assigned to each orientation. Fig. \ref{MC_texture} (a) shows the orientation-dependent Gaussian shape bulk energy:
\begin{equation}
b(n) =\frac{1}{(32\pi)^{0.5}}Exp[-\frac{(n-46)^{2}}{2\times46^{2}}]  ,
\label{flow}
\end{equation}
where $n$ is orientation, and $b$ is the bulk energy associated with this material. The bulk energy is adimensioned within a pure MC setting. MC simulations were carried out at an effective temperature $\alpha_{mc}=1.0$. Fig. \ref{MC_texture} (b) presents the orientation distribution at three time slices. As physically expected, materials points with a lower bulk energy are favored. This indicates that grains which experience the least plastic deformation will be favored as opposed to the ones have larger dislocation content.

The parallelized HMC approach was then used to consider the microstructural evolution of three-dimensional, polycrystalline nickel in response to plastic deformation. Attention was restricted to texture development in response to plastic deformation which is dominated by grain boundary kinetics rather than that of triple junction kinetics~\cite{Gottstein20051535,Gottstein20061065}. To meet this requirement, all numerical experiments were performed at an MC effective  temperature ($\alpha_{mc}=2.6$), which is a non-physical temperature. Note that the corresponding physical temperature (as opposed to MC temperature) is still not high enough to initiate recrystallization. In addition, elastic driving forces were suppressed for the sake of clarity. The MCP framework is implemented in MPM as a plastic constitutive model mediating the mechanical response~\cite{Miehe.2001}. The required fitting parameters are listed in Table \ref{plapara}. Material properties at room temperature were adopted even though they may be temperature dependent~\cite{Ledbetter.1982}. In order to illustrate the methodology and facilitate the interpretation of the results, grains were distinguished by a single angle of rotation with respect to the z-direction and allowed it to vary from $0^{\circ }$ to $90^{\circ }$ in $1^{\circ }$ angle steps. The other two Euler angles were held fixed. A similar setting has been previously consider with deformation restricted to the elastic regime~\cite{Zhang2010419}. 

\begin{figure}[ptb]\begin{center}
\includegraphics[width=0.45\textwidth]{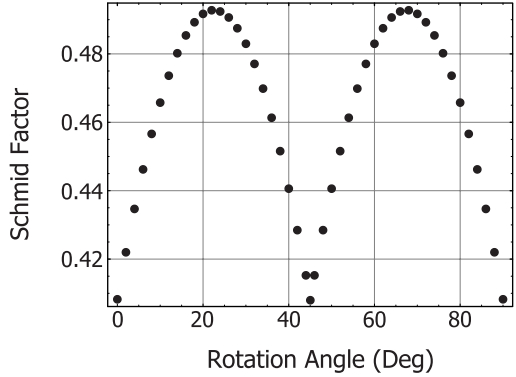}
\caption{Orientation-dependent Schmid factors. Each orientation is described by a single orientation angle with respect to the z-direction.}
\label{schmid_orn}
\end{center}\end{figure}

\begin{figure}[ptb]\begin{center}
\includegraphics[width=0.6\textwidth]{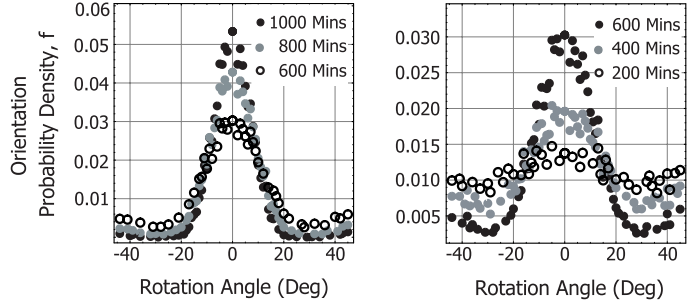}
\caption{Texture evolution under a uniaxial plastic loading in the x-direction. Each orientation is represented by a single Euler angle. Simulation data are shown as discrete points, which are averaged over $10$ independent runs.}
\label{texture_pla}
\end{center}\end{figure}

As mentioned previously, plastic slip occurs when the resolved shear stress on a slip system exceeds its threshold resistance. As a consequence, orientations with bigger Schmid factor will be plastically deformed first. Fig. \ref{schmid_orn} shows the orientation dependent Schmid factor of a face centered cubic (fcc) single crystal. Orientations near $22.5^{\circ}$ and $67.5^{\circ }$ have the largest Schmid factor. These orientations slip more easily and thus accumulate dislocations the fastest. Therefore, subsequent grain boundary motion will tend to remove such grains. In the current study, an $20$ MPa uniaxial loading was applied in the x-direction in $100$ loading increments. At each step, the strain increment associated with each MPM particle is computed using the MPM algorithm, and the MCP algorithm is then called to partition elastic and plastic deformations. The quasi-static state of the polycrystalline system is reached after a series of iterations when either the relative stress increment or relative strain increment meets the prescribed convergence criterion for all the particles~\cite{Zhang2010419}. After each MCP step, dislocation energy is computed and converted into the MC domain using our parametric links. The microstructure is subsequently evolved for one MC step, equal to $10$ minutes~\cite{Zhang2010419}.  Figs. \ref{texture_pla} (a) and (b) describe the evolution of orientation distribution at five different time slices. Negative rotation angles are used to represent orientation from the $46^{\circ }$ to $90^{\circ }$. As expected, texture evolution favors grains which have smaller Schmid factors with respect to the loading axis. 

\begin{figure}[ptb]\begin{center}
\includegraphics[width=0.6\textwidth]{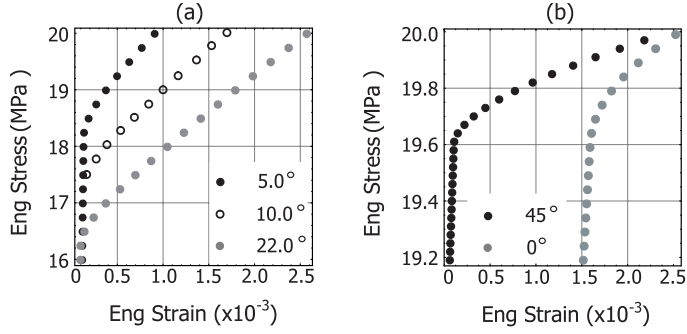}
\caption{Anisotropic behaviors of single crystals under plastic deformation: (a) stress-strain relationship for three single crystals with different Schmid factors, (b) stress-strain relationship for two crystals with the same Schmid factors.}
\label{orn_sgl}
\end{center}\end{figure}

Interestingly, orientations between $23^{\circ }$ and $45^{\circ }$ are not selected with the same frequency as orientations between $0^{\circ }$ and $22^{\circ }$ even though the two groups have identical Schmid factors. This is because the effective Young's moduli of the former group elements are larger than that of their counterparts. Therefore less elastic, and thus more plastic strain, results from the same loading. To illustrate this principle more clearly, pure mechanical simulations were performed in single crystals of various orientations. Fig. \ref{orn_sgl} (a) shows that more plastic deformation is obtained on the $22^{\circ }$ orientation than the other two orientations. Fig. \ref{orn_sgl} (b) highlights the differences in the mechanical response between $0^{\circ }$ and $45^{\circ }$ orientations which have the same Schmid factor. As expected, more plastic deformation is observed in the $45^{\circ }$ orientation.

\begin{figure}[ptb]\begin{center}
\includegraphics[width=0.6\textwidth]{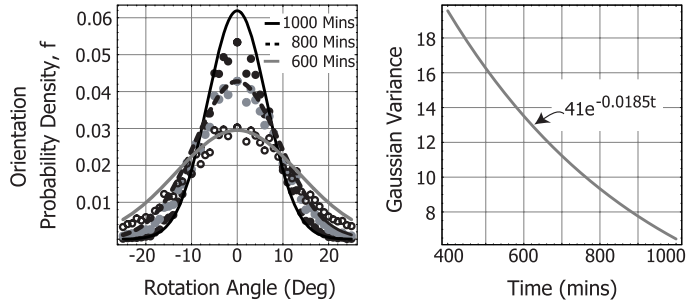}
\caption{Numerical fits of texture evolution with plastic influences: (a) Simulation data are shown as discrete points along with Gaussian fits as solid curves, and (b) Evolution of the Gaussian variance for the orientation distributions.}
\label{fitted_texture}
\end{center}\end{figure}

To quantify the effect of texture evolution with plasticity, the texture histograms were fitted to a time dependent Gaussian equation~\cite{Zhang2010419}:
\begin{equation}
f(\phi ,t)= \frac{1}{g(t)\sqrt{2\pi}}Exp[-\frac{\phi ^{2}}{2g(t)^{2}}],
\label{text_evl}
\end{equation}
where $f$ is the orientation probability density, $\phi$ is the orientation angle, and $g$ is a time-dependent Gaussian variance. The fitted results are shown as solid curves in Fig. \ref{fitted_texture} (a) along with the Gaussian variance described in Fig. \ref{fitted_texture} (b).

\section{Discussion and Summary}

The effects of plastic deformation on texture evolution are quantified with a previously developed hybrid algorithm that blends a discrete, probabilistic algorithm for grain boundary motion with a continuum, deterministic model of elastic deformation. Plastic deformation is accounted for using a Monte Carlo plasticity model in which slip occurs through a sequence of probabilistically determined, single slip events. Mechanical loading results in an inhomogeneous distribution of dislocation energy which amounts to a plastic driving force for texture evolution.  Grains with less damage grow at the expense of those which have been more heavily deformed. 

In the current approach, the dislocation density associated with the area swept by a moving interface is replaced with that of the overtaking grain. This implies that, when a grain boundary moves, the dislocation content adjacent to the grain boundary is extended into the material through which the grain boundary migrates. However, this is not always the case in reality. On the other hand, if one assume that no dislocations were transmitted with the boundary motion, then a new dislocation free grain would be introduced and the material would recrystallize~\cite{Humphreys.2004}. Though recrystallization undoubtedly occurs, and is certainly important in the materials studied and at these levels of deformation, it is not the only option. A grain boundary can simply migrate in a heavily deformed polycrystal~\cite{Beck.1952}. The migration of a grain boundary in a deformed polycrystal involves the propagation of some, but not always all, of the dislocation content that are adjacent to it. The authors are not aware of a meaningful quantitative description of how the dislocation structure propagates with a migrating grain boundary, and thus assume that nearly all the dislocations are carried with a boundary when it moves.

This hybrid computational methodology enables the temporal evolution of the microstructure under thermomechanical loading. It offers an alternative to sharp-interface and phase-field modeling. The resulting texture evolution maps are expected to be useful in materials and process design, thermomechanical fatigue as well as in part performance assessment throughout service life. The current work was limited to a single angle of misorientation between grains. An extension of this work to include fully arbitrary misorientation is underway, which will allow us to capture the influence of plastic deformation on the texture evolution of more realistic material systems.

\section{Acknowledgments}
The research is funded by Sandia National Laboratories. Sandia National  Laboratories are operated by the Sandia Corporation, a Lockheed Martin Company, for the United States Department of Energy under contract DE-AC04-94AL85000.  We also acknowledge the Golden Energy Computing Organization at the Colorado School of Mines for the use of resources acquired with financial assistance from the National Science Foundation and the National Renewable Energy Laboratories.

\section{references}

\end{document}